\documentclass[12pt,english]{iopart}
\usepackage{babel}
\usepackage{graphicx}
\usepackage[gray]{xcolor}
\definecolor{red}{gray}{0}

\usepackage{iopams}

\begin{document}

\title[Bang-bang shortcut to adiabaticity in the Dicke model as realized in a \ldots]{Bang-bang shortcut to adiabaticity in the Dicke model as realized
in a Penning trap experiment }

\author{J. Cohn$^{1}$, A. Safavi-Naini$^{2,3}$, R. J. Lewis-Swan$^{2,3}$,
J.G. Bohnet$^{4}$, M. G\"arttner$^{2,3,5}$,
K.A. Gilmore$^{4}$, E. Jordan$^{4}$, A. M. Rey$^{2,3}$, J. J. Bollinger$^{4}$,
J. K. Freericks$^{1}$}

\address{$^{1}$Department of Physics, Georgetown University,
Washington, DC 20057, USA}
\address{$^{2}$JILA, NIST and University of Colorado, 440
UCB, Boulder, CO 80309, USA}
\address{$^{3}$Center for Theory of Quantum Matter, University
of Colorado, Boulder, Colorado 80309, USA}
\address{$^{4}$National Institute of Standards and Technology, Boulder, Colorado 80305, USA}
\address{$^{5}$Kirchhof-Institut
f\"ur Physik, Universit\"at Heidelberg, Im Neuenheimer
Feld 227, 69120 Heidelberg, Germany}
\begin{abstract}
We introduce a bang-bang shortcut to adiabaticity for the Dicke model, which we implement via a 2-D array of trapped ions in a Penning trap with a spin-dependent force detuned close to the center-of-mass drumhead mode. Our focus is on employing this shortcut to create highly entangled states that can be used in high-precision metrology. We highlight that the performance of the bang-bang approach is comparable to standard preparation methods, but can be applied over a much shorter time frame. We compare these theoretical ideas with experimental data which serve as a first step towards realizing this theoretical procedure for generating multi-partite entanglement.

{\it This manuscript contains contributions from NIST, and is not subject to U.S. copyright.}
\end{abstract}

\maketitle

\section{Introduction}

The field of quantum metrology has the potential to drastically improve precision measurements from the standard quantum limit to the Heisenberg limit. These techniques rely on the ability to create entangled quantum states and employ them, via interferometric methods, to produce high-accuracy measurements. A range of different techniques can be employed to harness the metrological applications of a variety of entangled states~\cite{gilchrist2004schrodinger,giovannetti2011advances,leibfried2005creation,strobel2014fisher,biercuk2010ultrasensitive}. 

Creating these metrologically useful states is generally a difficult task. One promising method is adiabatic state preparation, where the system starts with a simple Hamiltonian that has an easily produced product state as its ground state and is then adiabatically evolved to the entangled ground state of a complex Hamiltonian by slowly varying an external parameter. The challenge is that the adiabatic state preparation must be done slowly compared to the relevant minimum energy gap to reduce unwanted diabatic excitations during the evolution. For systems that have vanishing gaps in the thermodynamic limit, the minimal gap for a finite system often decreases inversely with the system size making adiabatic state preparation particularly difficult for larger systems. Current quantum simulators cannot evolve the system long enough to be able to fully carry out this process, as they are limited by decoherence and technical noise. This constraint, of a short evolution time, inevitably produces diabatic excitations, which can be significant and can seriously affect the fidelity of the target entangled state. The challenge lies in finding balance between decoherence errors entering on long time scales and the diabatic excitations entering on short time scales.     
  
One potential solution to this problem is a shortcut to adiabaticity---the system is evolved non-adiabatically so that it ends up in the entangled ground state at the end of the evolution. These techniques reduce the total state preparation time, which make them attractive when dealing with decoherence effects. Lately, there have been many theoretical breakthroughs in this
area~\cite{demirplak2005assisted,deffner2014classical,berry2009transitionless}.
One technique, based on adding counter-diabatic fields to the Hamiltonian, guarantees that the system evolves to the correct entangled ground state. It does this by adding an auxiliary term to the Hamiltonian, which is designed to exactly cancel the excitations that would take place, ensuring that the system always remains in the instantaneous ground state. The strength of this term goes to zero at the end of the ramp, which results with the system in the entangled ground state of the target Hamiltonian. Unfortunately the auxiliary terms that must be employed require a large number of nonlocal and time-dependent interactions to be added to the Hamiltonian,
which are difficult to implement due to their complexity. Recent advances~\cite{sels2016minimizing} show that while exact counter-diabatic driving may not be realized in real systems, local counter-diabatic terms may be applied to reduce the diabatic excitations. These techniques would increase the ground-state fidelity, but for this work the terms used to construct them generally break a parity symmetry that protects the entangled states. 

An alternative approach is to try to minimize the diabatic excitations by ramping quickly when the instantaneous energy gap is large and slowly when it is small, given the constraint of the total experimental run time. 
Implementing this logic continuously results in a ramping scheme termed the
locally adiabatic (LA) ramp~\cite{richerme2013experimental}. Here,
the ramp speed for the external parameter is optimized by ensuring the diabatic excitations are created at a uniform rate throughout the ramp. It requires knowing the instantaneous
minimum energy gap within the same symmetry sector as the ground-state, so it is challenging to implement for systems where this gap is not known {\it a priori}. There is a conjecture that this is the best continuous ramp profile to use for a given experiment if the energy gap in the given symmetry sector is known and the experimental run time is long enough to achieve reasonable fidelity~\cite{richerme2013experimental}.

The bang-bang protocol~\cite{viola1998dynamical,balasubramanian2015bang}, 
presented here, is a more widely applicable alternative, because it does not require one to know the minimal energy gap as a function of time. It consists of (i) initializing the system in a convenient product state (usually chosen to be the ground-state
of the initial ``simple'' Hamiltonian); (ii) quenching the external parameter to an intermediate Hamiltonian (which often has a gap close to the minimal gap of the system) and holding for a period of time and (iii) then quenching the external parameter to the final Hamiltonian of interest.
The procedure involves
optimizing two parameters: the external parameter for the intermediate Hamiltonian
and the holding time. {\color{red} In earlier work, the protocol was shown to work better for longer-range interactions~\cite{balasubramanian2015bang}.}

{\color{red}
In this work, we experimentally implement the bang-bang and LA protocols in a system of $\sim 70$ trapped Be$^+$ ions forming a two dimensional (2-D) planar Coulomb crystal.
The trapped-ion system realizes a quantum simulator of the Dicke model, which describes the behavior of a large collective spin coupled to a single radiation mode in the presence
of an additional transverse field coherently driving the spin~\cite{dicke1954coherence, safavi2017exploring}; here the radiation mode is the center-of-mass phonon mode. The model possesses a quantum critical point separating two distinct quantum phases:  the superradiant phase  characterized by a macroscopic population of the radiation mode and  ferromagnetic spin correlations  and the
normal phase where the radiation field remains in vacuum and  the spins are aligned to the  strong external field. We investigate the performance of each protocol when preparing the ground-state of the Dicke model in
the superradiant phase,  which is a multi-partite entangled state  optimal  for quantum sensing protocols~\cite{toscano2006sub}. We experimentally characterize the performance of each protocol using collective spin observables and
full spin distribution functions and compare them to extensive theory calculations. The latter  also allow us to benchmark the performance of the protocols based on  ground-state fidelity and quantum Fisher information.
In Sec.~II we first outline the Dicke model, following Ref.~\cite{safavi2017exploring}. We present experimental observations for the implemented ramps and accompanying theoretical calculations in Sec.~III.  In Sec.~IV
we discuss how the protocols may be optimized for the production of metrologically useful entangled states. Lastly, in Sec.~V we make concluding remarks.
}

\section{Formalism {\color{red} and Description of the System}\label{sec:formalism}}

{\color{red} We consider a  trapped-ion system of laser-cooled  $^9$Be$^{+}$ ions in a Penning trap. The interplay of the Coulomb repulsion and the external electromagnetic trapping
potentials stabilizes a 2-D planar crystal. The valence electron spin states in the ground state of the ion encode the spin-one-half degree of freedom, while the normal vibrational modes of
the self-assembled Coulomb crystal form the bosonic degree of freedom (phonons). In the 4.46 T magnetic field of the Penning trap, the electronic states are split by 124 GHz.
A pair of laser beams couple  the spin and phonon degrees of freedom. By adjusting the  detuning of the lasers close to the center of mass (COM) mode of the crystal, only this  mode is excited and the spin-phonon coupling becomes uniform throughout the
system. In this regime, the experimental system can be described by the Dicke Hamiltonian, defined to be $\mathcal{H}_{Dicke}(t)=\mathcal{H}_{Ph}+\mathcal{H}_{int}(t)+\mathcal{H}_B(t)$, with}
\begin{eqnarray}
\mathcal{H}_{Ph}&=&\omega_{COM}\hat{a}^{\dagger}_{COM}\hat{a}_{COM}, \label{eq: ham_PH} \\
\mathcal{H}_{int}(t)&=&-\frac{2g}{\sqrt{N}}(\hat{a}_{COM}+\hat{a}_{COM}^{\dagger})\hat{S}_{z}\cos(\mu t), \label{eq: ham_int} \\
\mathcal{H}_{B}(t)&=&B^{x}(t)\hat{S}_{x}. \label{eq: ham_B}
\end{eqnarray}
Here $\hat{a}_{COM}$ ($\hat{a}^{\dagger}_{COM}$) are the phonon annihilation (creation) operators for the COM mode with frequency of $\omega_{COM}$ ($[\hat a_{COM},\hat a^\dagger_{COM}]=1$), $g$ is the spin-phonon coupling strength, $\mu$ is the {\color{red} beat-note frequency of the Raman lasers driving the system} and $B^{x}(t)$ is the time-dependent transverse field {\color{red}  (we work in units with $\hbar=1$ and $g_{eff}\mu_B=1$). As the coupling is uniform, the spin degree of freedom is described by collective operators where $\hat{S}_{\alpha}=\sum_{i}\hat{\sigma}^{\alpha}_{i}$ and
$\hat{\sigma}_i^\alpha$ is the Pauli spin matrix at site $i$ with $\alpha = x, y, z$ ($[\sigma_j^\alpha,\sigma_k^\beta]=i\delta_{jk}\epsilon_{\alpha\beta\gamma}\sigma^\gamma_j$). Moreover, as the Dicke Hamiltonian conserves the total spin, we may restrict our Hilbert space to the
$N+1$ Dicke states that span  the maximal spin multiplet   (since the ground-state is the global ground-state over all possible multiplets), enabling us to numerically simulate the quantum dynamics of large systems.}
 When the transverse field goes to zero, $\hat{S}_z$ commutes with the Hamiltonian and the spin components of the eigenstates take the form of $\hat{S}_z$ projections within the maximal spin multiplet (subject to possible degeneracies of different spin projections).       

{\color{red} Our calculations are facilitated further by implementing the rotating wave approximation (RWA) within the frame rotating with an angular velocity $\mu$. In this frame, we recover the Dicke Hamiltonian given by
\begin{equation}
\label{eq:Dicke}
\mathcal{H}_{Dicke}^{RWA}(t)=-\delta\hat{a}_{COM}^{\dagger}\hat{a}_{COM}-\frac{g}{\sqrt{N}}\left(\hat{a}_{COM}^{\dagger}+\hat{a}_{COM}\right)\hat{S}_{z}+B^{x}\left(t\right)\hat{S}_{x},
\end{equation}
with $\delta=\mu-\omega_{COM}$.
We always have $\delta<0$, so that the first term in the Hamiltonian is positive.} Note that the $z$ and $x$ projections are interchanged from the standard form of the Dicke Hamiltonian~\cite{garraway2011dicke}.

{\color{red} While it is not possible to find an analytic expression for the ground-state of the Dicke Hamiltonian generally, it is possible to do so in certain regimes.
We begin by rewriting the Hamiltonian in Eq.~(\ref{eq:Dicke}) as
\begin{equation}
 \mathcal{H}_{Dicke}^{RWA}(t) = -\delta \hat{b}^{\dagger}\hat{b} + \frac{g^2}{N\delta} \hat{S}_z^2 + 2B^x(t)\hat{S}_x, \label{eqn:SpinBoson_H}
\end{equation}
where $\hat{b} = \hat{a}_{COM} - (g/\sqrt{N}\delta)\hat{S}_z$. In this form, the ground-state can be well understood in two distinct regimes: the weak-field $B^x \ll g^2/4|\delta|$ (superradiant) and
strong-field $B^x \gg g^2/4|\delta|$ (normal) limits. A quantum critical point separates these phases at $B^x \sim g^2/4\vert \delta\vert$.

In the weak-field limit, $B^x \ll g^2/4|\delta|$, the energy of the Hamiltonian is minimized by aligning all spins along $\pm \hat{e}_z$ and coherently displacing the phonons via the spin-dependent
displacement of $\pm\alpha\approx(g/\sqrt{N}\vert\delta\vert) \hat{S}_z$. This leads to a cat-like spin-phonon ground-state
\begin{equation}
 \vert \psi,B^x\to 0\rangle = |CAT(\alpha)\rangle \equiv |\alpha,0\rangle_{ph}\otimes|+N/2\rangle_z + |-\alpha,0\rangle_{ph}\otimes|-N/2\rangle_z,
 \end{equation}
where $|\alpha,n\rangle=\hat{D}(\alpha)|n\rangle$ is a displaced Fock state and $\hat{D}(\alpha) = \exp({\alpha\hat{a}_{COM}^{\dagger} - \alpha^*\hat{a}_{COM}})$ is the displacement operator.

Conversely, in the strong-field limit, $B^x \gg g^2/4|\delta|$, the nature of the ground-state will be dominated by the transverse field and is characterised as all spins aligned against the
field, i.e. pointing along  $ -\hat{e}_x$ (here we assume $B^x>0$ for simplicity). Given this spin-orientation, the displacement of the phonons vanishes and the  spin-phonon ground-state is
\begin{equation}
 \vert \psi, B^x\to\infty\rangle = \vert0\rangle\otimes\vert -N_s/2 \rangle_x .
\end{equation} A schematic representation of different low-energy eigenstates
is given in  Fig.~\ref{fig: phases} for the phonon-like regime.

\begin{figure}[h]
	\centerline{
		\includegraphics[scale=0.85]{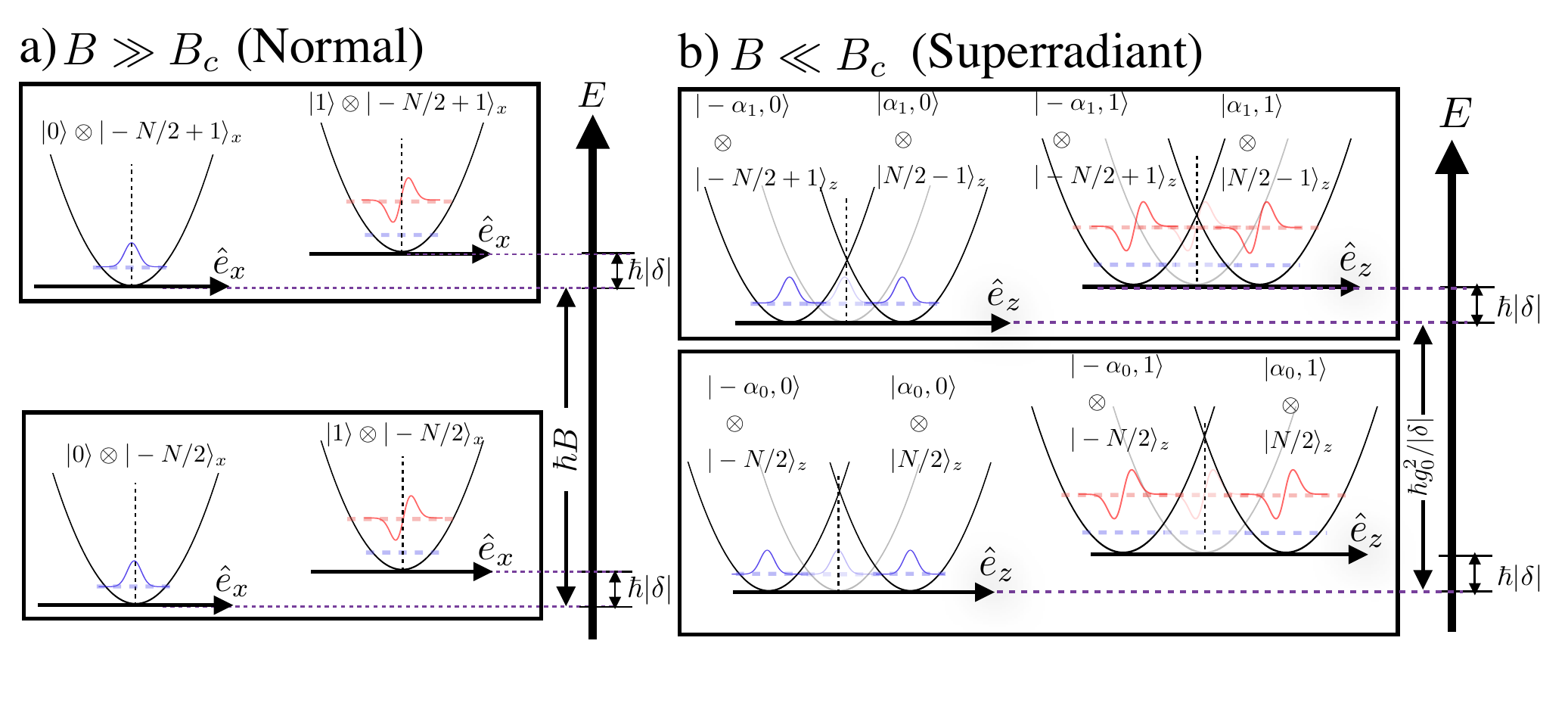}
	}
	\caption{Schematic diagram of the ground states of the Dicke model for the normal and superradiant phases.   (a) The energy eigenstates of the normal phase are represented by the phonon Fock states, $|n\rangle$, and the spins oriented along the $x$-axis. If $B^{x}>\delta$ {\color{red}(plotted here)}, the low lying excitations are phonon like, and if $B^{x}<\delta$ {\color{red}  (not shown)} they are represented by spin flips along the $x$-axis. (b) The energy eigenstates in the superradiant phase, where the phonons are represented by displaced Fock states, $\hat{D}(\alpha)|n\rangle$, and the spins are aligned in the $\pm z$-direction. In this region, the low lying excitations are phonon like if $g^{2}/\delta > \delta$ {\color{red} (plotted here)} and are represented by spin flips along the $z$-axis if $g^{2}/\delta < \delta$ {\color{red} (not shown)}.    {\color{red} The symbol $\hat e_i$ denotes the unit vector in the $i$ direction.}   \label{fig: phases} }
\end{figure}

The capability of a ramping protocol to satisfy the adiabatic condition in our  system  is intimately determined by the energy gap of the Dicke Hamiltonian. This is particularly relevant for  the LA protocol which,
as we will detail in the following section, requires full knowledge of the energy gap. We note also that while the bang-bang protocol isn't a smooth ramp but rather a double quench,
we still expect that a smaller gap will generate more unwanted excitations following the quench, thus reducing the efficiency of ground-state production.
For the Dicke Hamiltonian, the size of the gap generally depends on $\delta$ in a complex manner as  discussed in detail in Ref.~\cite{safavi2017exploring}. However, in qualitative terms
the gap increases with detuning $\delta$, as long as we can keep the effective coupling strength $g^2/|\delta|$ fixed.}

{\color{red}
While the size of the energy gap can be problematic for adiabaticity, the Dicke model also possesses symmetries which increase the efficiency of ground-state preparation. Specifically, the Dicke Hamiltonian
is symmetric with respect to the transformation of the spin operators $\hat{S}_x\rightarrow\hat{S_x}$, $\hat{S}_y\rightarrow -\hat{S}_y$, and $\hat{S}_z\rightarrow -\hat{S}_z$ (this is equivalent to a $\pi$ rotation of the spins
about the $x$-axis), and a transformation of the phonon momentum and position operators ($\hat{p}\rightarrow -\hat{p}$ and $\hat{x}\rightarrow -\hat{x}$),
or equivalently the raising and lowering ($\hat{a}^\dagger_{COM}\rightarrow -\hat{a}^\dagger_{COM}$ and $\hat{a}_{COM}\rightarrow -\hat{a}_{COM}$).
This symmetry allows us to characterize the eigenstates as even or odd parity under the spin reflection operation
(when expressed in the $z$ or $y$ spin bases) plus an inversion of the phonon coordinates, with associated conserved quantity
$\langle\exp[-i\pi(\hat{n}_{COM}+\hat{S}_x)]\rangle$. This symmetry restricts the available Hilbert space to states with the same parity. 
More explicitly, if the system is initialized in the ground state at large
$B^x$ ($|0\rangle_{ph}\otimes|-N/2\rangle_x$), then states are restricted to the even parity sector if $N$ is even, and restricted to the odd parity sector if $N$ is odd during the ramp.
This implies that the relevant gap to determine the rate of diabatic excitations is the energy gap to the first excited state \textit{in the same symmetry sector as the ground state}. In the presence of diabatic excitations, this enlarged energy gap helps maintain multipartite entanglement and metrological utility in the final state.
Note that if a longitudinal magnetic field (in the $z$-direction) is added to the Dicke model, breaking the spatial-spin reflection parity symmetry, this can rapidly lead to a degradation of
the entanglement in the system. In the experiment, stray longitudinal fields do occur and will need to be controlled in order to achieve optimal cat-state production~\cite{safavi2017exploring}.}

\section{Experimental Results}

{\color{red} We now present a comparison between the experimental observations and theoretical simulations for the bang-bang and LA protocols.
The theoretical simulations were carried out by time evolving the total quantum state while assuming perfect state preparation, spin operations,
and measurement readout. We note that despite the imperfections in the experiment, computational complexity prevents us from fully incorporating decoherence into the theoretical simulations.

The experimental sequence uses resonant 124 GHz microwave pulses to create
arbitrary collective spin rotations. These allow the initial state to be completely polarized
along the $x$-axis. Resonant microwaves are also used to generate
the transverse field. Projective collective spin measurements
are performed at various times by first rotating the desired spin axis
to the $z$-axis and then using global ion fluorescence to image the spin states (the up spins are bright and the down spins are dark).

The experiment was operated at $g=2\pi\times 0.935$~kHz and at a detuning of $\delta=-2\pi\times1$~kHz from the COM mode, where the spins and the phonon model are uniformly coupled and the RWA is valid. The initial transverse field was set to $B^{x}(t=0)=2\pi\times7$~kHz.
We note that the proximity of $\delta$ to the critical point at $B_c$ makes ground-state preparation much more difficult, as discussed in Sec.~\ref{sec:formalism} ~\cite{safavi2017exploring}. However experimental considerations, in particular current decoherence rates, restrict us to operate the experiment in  this parameter regime.}

\begin{figure}[h]
	\centerline{
		\includegraphics[scale=0.6]{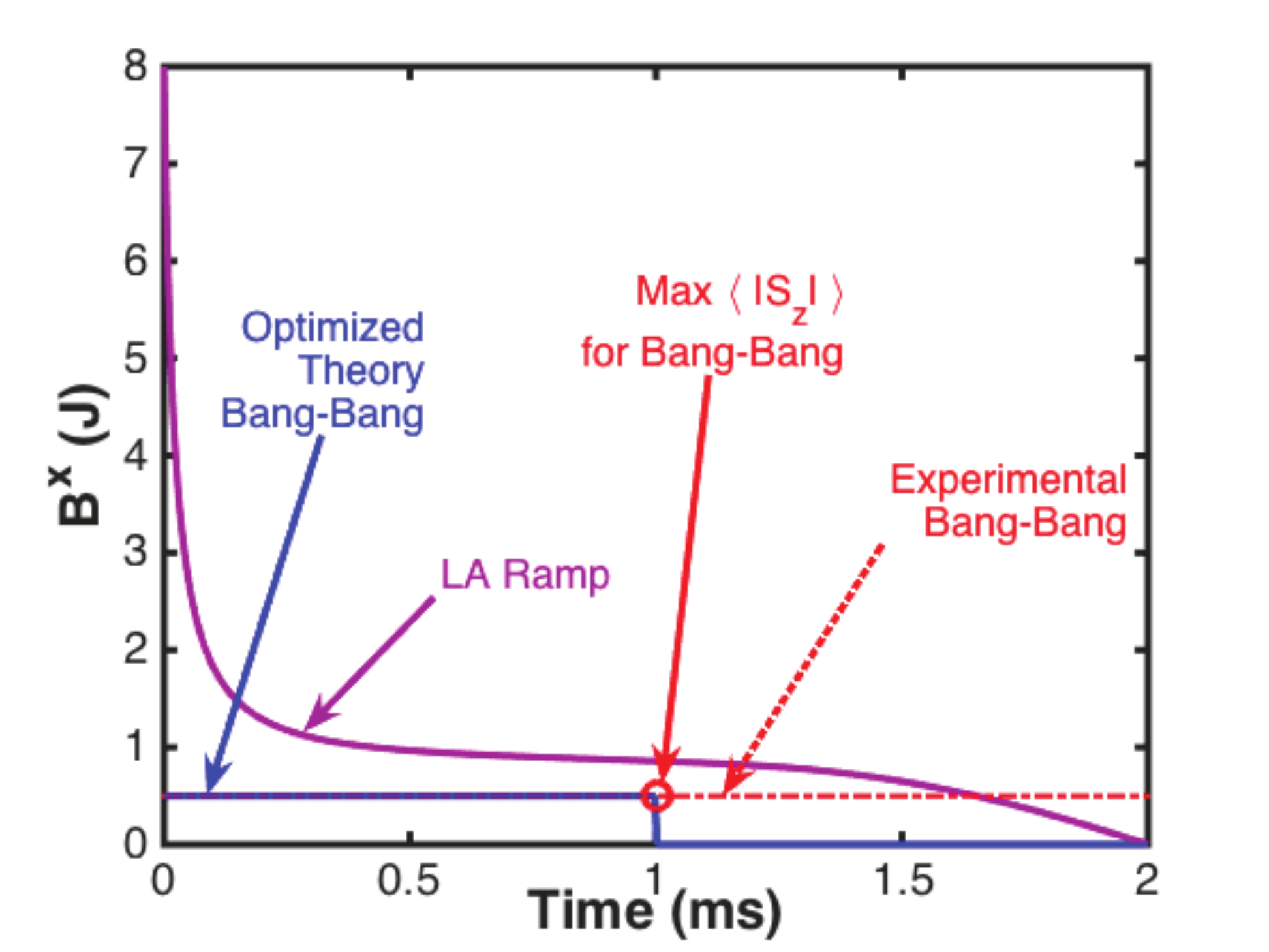}
	}
	\caption{Ramp profiles for the time-dependent transverse field in the Dicke model. We show the LA ramp and the bang-bang ramp. The LA and bang-bang ramps have been optimized to produce the highest ground-state fidelity for a simulation time less than or equal to 2 ms. {\color{red} The theoretical and experimental bang-bang ramps are optimized at about 1~ms (open circle). The experimental data was sampled out to 2 ms  with the same quench field.}  \label{fig: ramps} }
\end{figure}

{\color{red}
The experimental sequence was as follows: The initial state was prepared with all spins aligned along the $x$-axis. In the case of the bang-bang protocol, this was followed by a quench to an
intermediate transverse field. This intermediate quench was optimized in the lab to give the spins the largest possible projection onto the $z$-axis.
Note that, as shown in Fig.~\ref{fig: ramps}, the transverse field was not quenched to zero when the peak magnetization along the $z$-axis was reached.

The LA ramp profile was implemented according to the equation $\dot{B}(t) = \Delta(t)^2/\gamma$, for
\begin{equation}
 \gamma = \frac{\tau_{\mathrm{ramp}}}{\int_0^{B^x(0)}~ dB \frac{1}{\Delta(B)^{2}}} ,
\end{equation}
where $\Delta(t)$ [$\Delta(B)$] is the energy gap of the instantaneous Hamiltonian (at instantaneous field strength) and $\tau_{\mathrm{ramp}}$ is the total ramp duration. Essentially, the LA profile 
ramps the transverse field rapidly when the gap is relatively large, and is slowest when the gap reaches a minimum. 
Further discussion of the ramp and corresponding details of the experimental optimization procedure can be  found in Ref.~\cite{safavi2017exploring}.

In the absence of decoherence, we expect the final state to be the spin-phonon cat state (modified by the fact that the initial phonon population has $\bar n=6$), but we cannot tell whether such a state was actually formed from our data because we only measured the spin properties. We did not measure the spin-phonon entanglement.

Typical examples of the bang-bang and LA ramps are shown in Fig.~\ref{fig: ramps}. We note that the bang-bang ramp is similar to an extreme limiting case of the
LA protocol, where the field is held constant near the critical point during the ramp.}

\begin{figure}
	\begin{centering}
		\begin{tabular}{|c|c|}
			\hline 
\tabularnewline
			a.\includegraphics[scale=0.4]{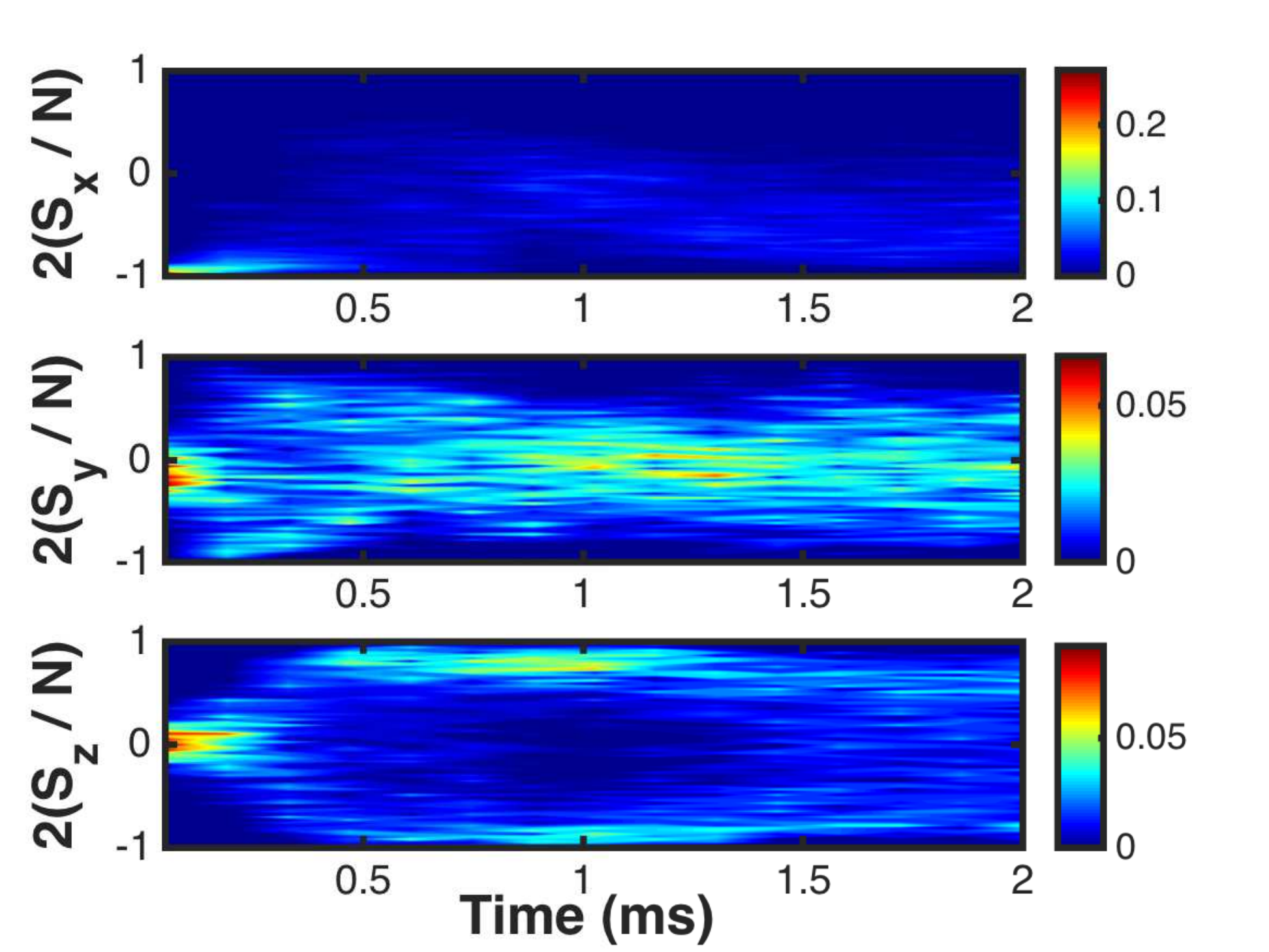} & b.\includegraphics[scale=0.4]{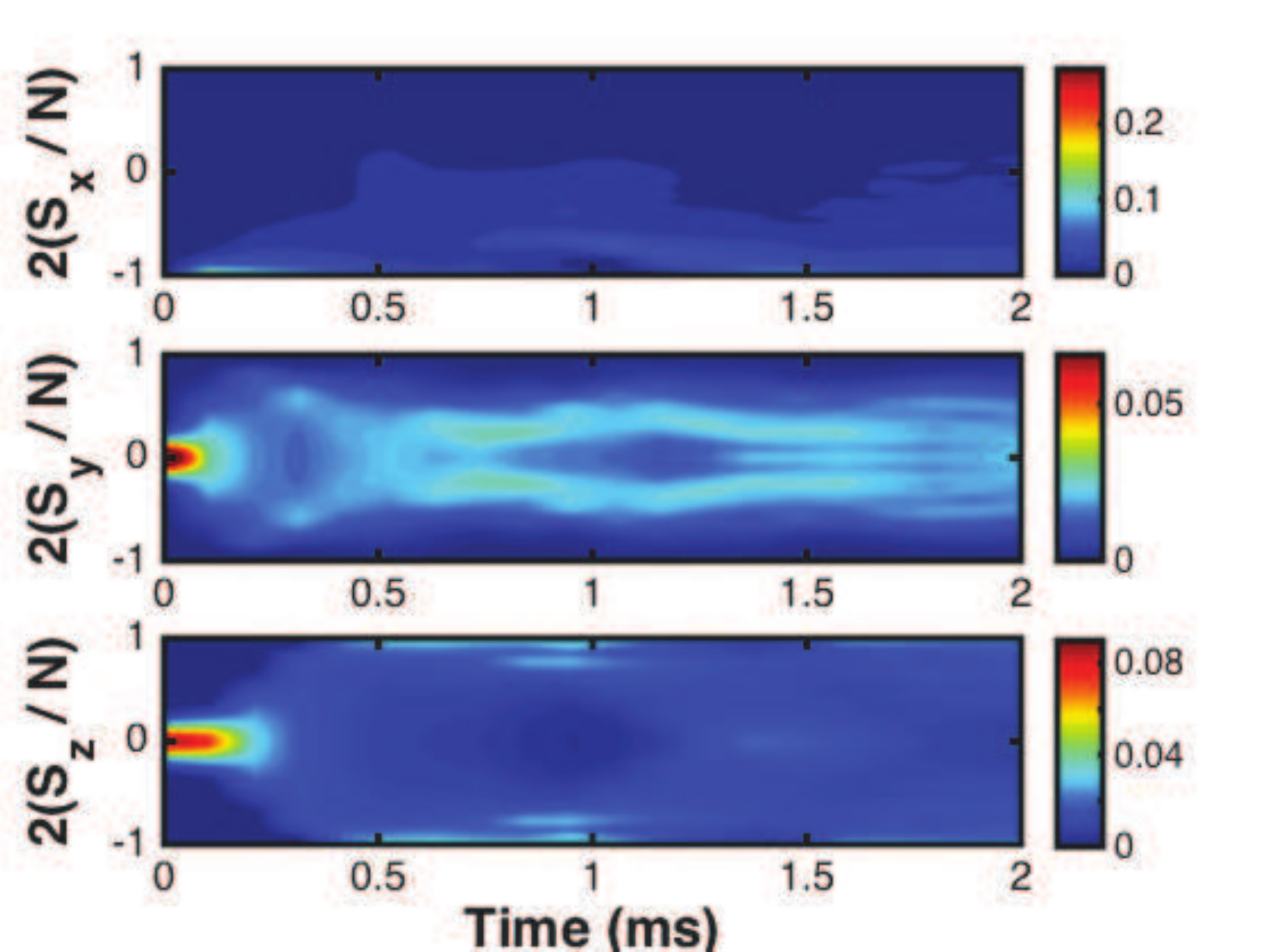}\tabularnewline
			\hline 
			\hline 
			c.\includegraphics[scale=0.4]{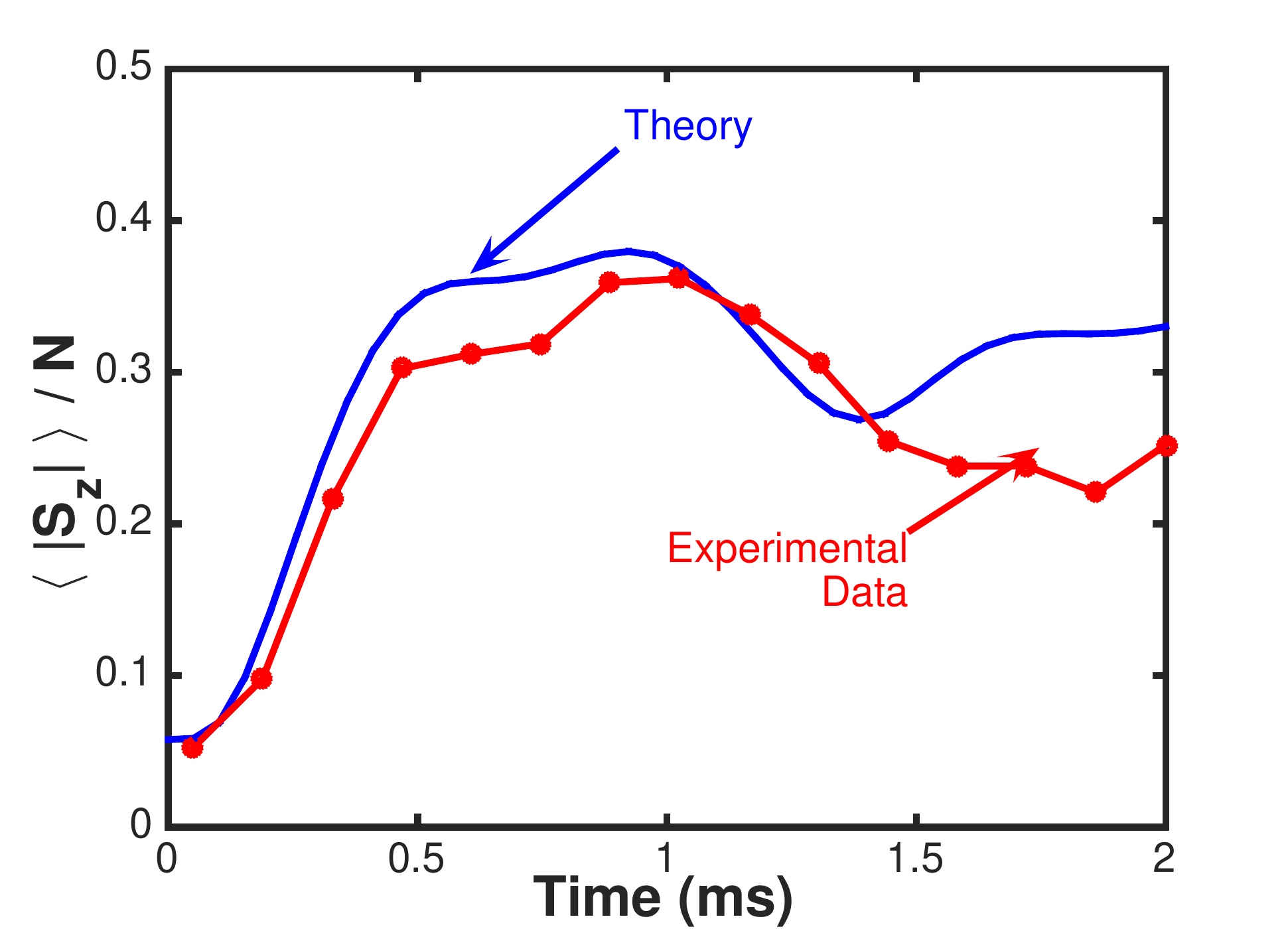} & d.\includegraphics[scale=0.4]{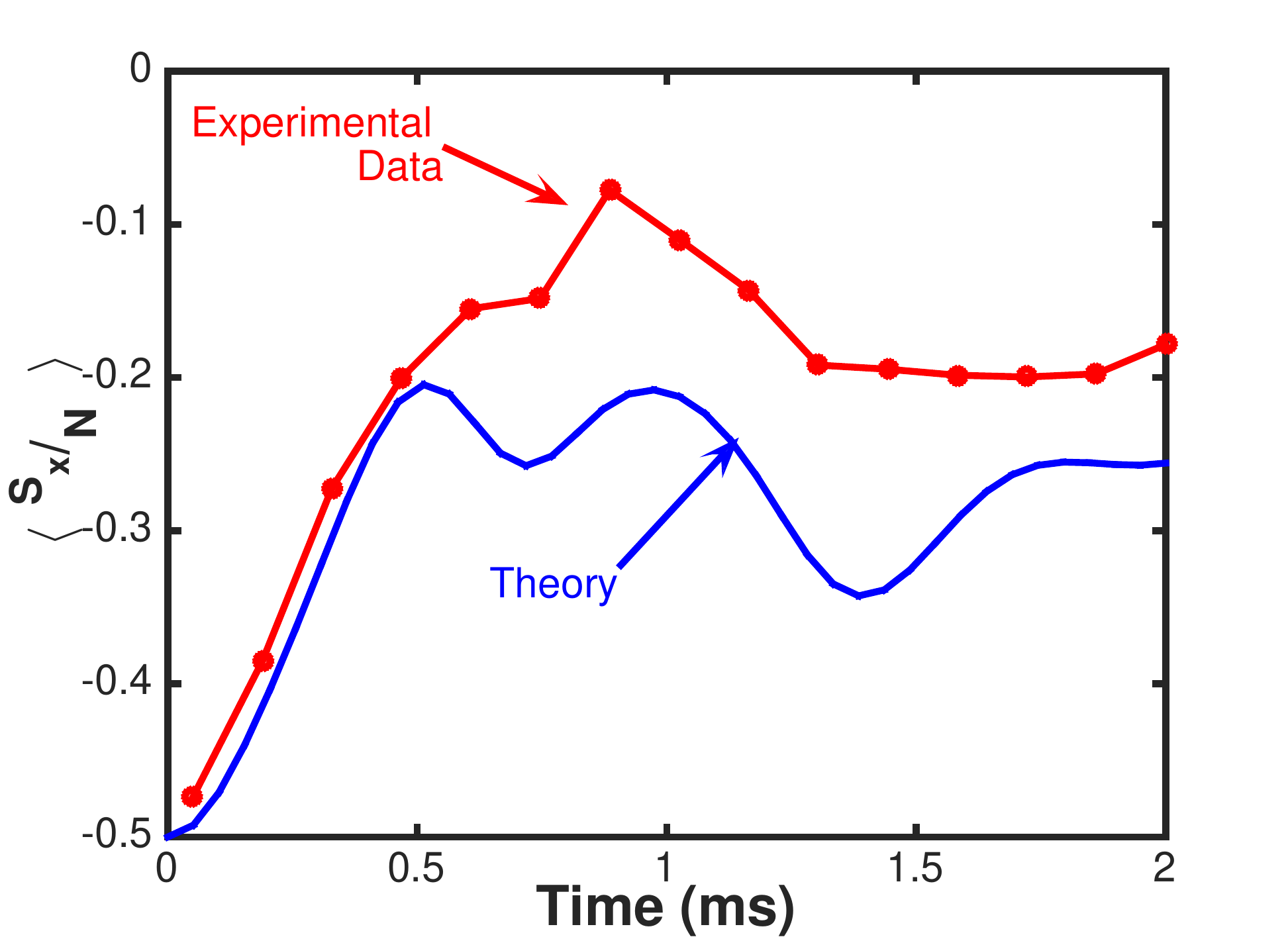}\tabularnewline
			\hline 
		\end{tabular}
		\par\end{centering}
	
	\caption{Comparison of experimental data and theory estimates for the optimal quench of the bang-bang ramp
		for a system of 75 ions with coupling constant $J=2\pi\times 0.875$~kHz,  and
		detuning from the COM mode of $\delta=-2\pi\times1$~kHz. The spins
		are initialized to the state $|-N/2\rangle_{x}$ and the
		COM mode is in a thermal state with an initial occupation of $\bar{n}\approx 6$. Figures (a) and
		(b) show plots of the experiment and theory respectively
		for the total spin projections in the $x$, $y$, and $z$ directions. Figure (c)
		shows the mean value of $\langle|S_{z}|\rangle/N$. {\color{red} A noticeable growth of $\langle|S_{z}|\rangle$ is observed after the initial quench. Figure (d) shows the mean value of  $\langle S_{x}\rangle/N$ which exhibits   fast  demagnetization. For this observable, however, dephasing plays a non-negligible role and the disagreement between theory and experiment becomes larger. The statistical error bars are on the order of the size of the data points.} \label{fig: expt}
	}

\end{figure}

\begin{figure}
	\begin{centering}
		\begin{tabular}{|c|c|}
			\hline 
\tabularnewline
			a.\includegraphics[scale=0.4]{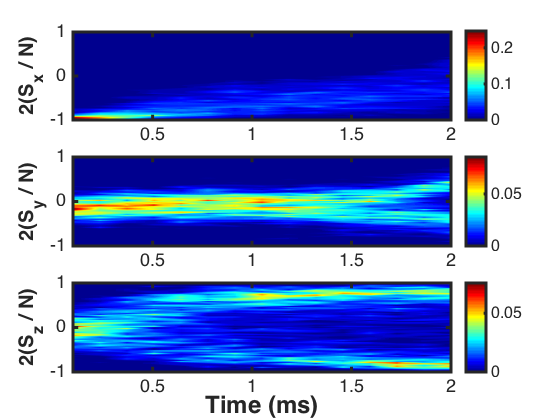} & b.\includegraphics[scale=0.4]{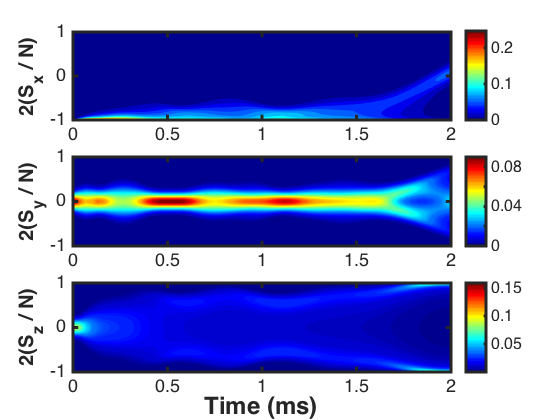}\tabularnewline
			\hline 
			\hline 
			c.\includegraphics[scale=0.4]{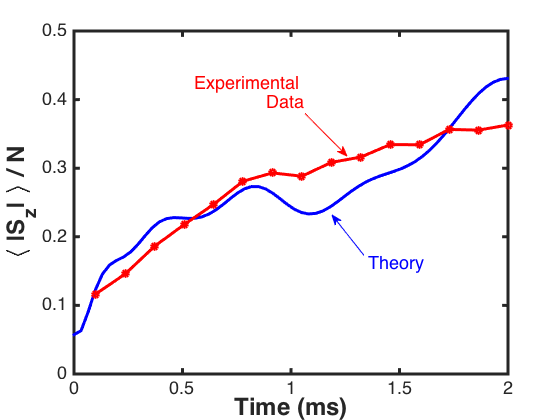} & d.\includegraphics[scale=0.4]{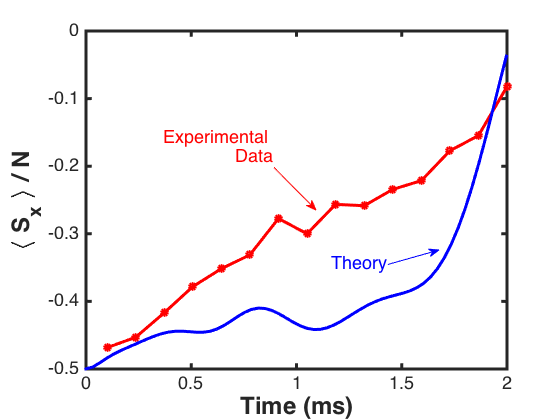}\tabularnewline
			\hline 
		\end{tabular}
		\par\end{centering}
	
	\caption{Comparison of experimental data and theory estimates for the LA ramp
		for a system of 76 ions with coupling constant $J=2\pi\times 0.875$~kHz,  and
		detuning from the COM mode of $\delta=-2\pi\times1$~kHz. The spins
		are initialized to the state $|-N/2\rangle_{x}$ and the
		COM mode is in a thermal state with an initial occupation of $\bar{n}\approx 6$. Figures (a) and
		(b) represent false-color plots of the experiment and theory respectively
		for the total spin projections in the $x$, $y$, and $z$ directions. Both of
		the $S_{z}$ plots show {\color{red} good} qualitative agreement. Figure (c)
		shows the values of $\langle|S_{z}|\rangle/N$.  {\color{red} A noticeable growth of $\langle|S_{z}|\rangle$ is observed in the superradiant regime. Figure (d) shows the mean value of  $\langle S_{x}\rangle/N$ which exhibits   fast  demagnetization. Similar to the bang-bang case in this observable dephasing plays a non-negligible role and the disagreement between theory and experiment becomes larger. The statistical error bars are on the order of the size of the data points.} \label{fig: la_exp}
	}

\end{figure}
          
The spin-projection plots in Figs.~\ref{fig: expt}~(a) and (b) show good qualitative agreement.
Both the experimental and theoretical data show an optimal peak in the
$S_{z}$-projection around 1 ms when the transverse field is initially quenched to $B^{x}=0.4\ J$.
A more detailed comparison can be seen in Figs.~\ref{fig: expt}~(c) and (d) where $\langle|S_{z}|\rangle$
and $\langle S_{x}\rangle$ are plotted as functions of time. 
These plots give close agreement between experiment and theory for times $t<$~0.4 ms, but even at longer times there is good qualitative
agreement.

{\color{red} In the case of the LA ramp, the qualitative behavior of the experimental data matches what
is expected by the theory, as shown in Figs~\ref{fig: la_exp}~(a) and (b). Figs~\ref{fig: la_exp}~(d) and (c) show $\langle|\hat{S}_z|\rangle/N$
and $\langle\hat{S}_x\rangle/N$ as a function of ramp time. Here, the LA ramp achieves a slightly larger $\langle|\hat{S}_z|\rangle$  at the end of the
$2$~ms ramp than the bang-bang data reaches at 1~ms, as expected. Fig.~\ref{fig: la_exp}~(d) shows a deviation of experimental and theory
plots of $\langle\hat{S}_x\rangle$ at short times, which hints that certain decoherence processes may also be present.}

Although the theory provides a qualitative understanding of the experimental results, there are clearly dynamics taking place which are not solely described by pure evolution under the Dicke Hamiltonian. {\color{red} We expect that} decoherence effects are the main contributor to this discrepancy. The two main sources of decoherence present in the experiment are Rayleigh and Raman scattering~\cite{safavi2017exploring,uys2010decoherence}. Rayleigh scattering causes the off-diagonal elements of the density matrix to be damped in the $\hat{S}_{z}$-basis, an effect also know as dephasing. Raman scattering produces spontaneous emission and absorption. Hence, Rayleigh scattering is {\color{red}expected to be} the main source of decoherence in these experiments~\cite{uys2010decoherence}. The dynamics of the density matrix is dictated by a master equation that satisfies, $\dot{\rho}=i[\hat{H},\rho]-\Gamma\sum_{i}(\rho-\hat{\sigma}^{z}_{i}\rho\hat{\sigma}^{z}_{i})$, where $\Gamma$ is the single particle decoherence rate due to Rayleigh scattering (measured in the lab to be $60s^{-1}$ at $B^x=0$).

{\color{red} While including the effects of decoherence along with the phonons and spins is too computationally costly for the system sizes considered in the experiment,
in certain limits, one can create phenomenological models for the effects of decoherence. In particular, when $B^x=0$, the coherences between different spin sectors,
$|m_i\rangle\langle m_j|$, in the density matrix decay as $\exp({-(m_{i}-m_{j})\Gamma t})$ where $m_{i}$ is a given eigenvalue of $\hat{S}_z$ \cite{garttner2017measuring}.
Unfortunately, this means that the coherence of an ideal spin-phonon cat state will decay exponentially with a rate that increases with ion number  since $m=|N/2|$. In the opposite regime, we attribute the rapid depolarization
of $\hat{S}_x$ at short times, in the presence of a dominant transverse field and for a state along the $x$-axis ($\langle \hat{S}_x\rangle\rightarrow\langle \hat{S}_x\rangle\exp[-\Gamma t]$), to
decoherence. We note that this condition is not present in the bang-bang experiment as the system is never in the large $B^x$ regime. If $B^x\sim|\delta|$, we are unable to develop a phenomenological
model for the effects of decoherence. However one expects that decoherence will still result in a reduced final magnetization. We have found that a generically  longer ramp time correlates to a
larger discrepancy between the experimental data and the theory estimates of $\langle|\hat{S}_z|\rangle$.}

The experiment did not attempt to disentangle the expected spin-phonon entanglement and transfer it to a spin-only entanglement, nor did it directly measure the entanglement of the final state. These are generally complex tasks which will be pursued in more detail in future experiments. Nevertheless this spin projection data does serve as a first step in understanding the evolution and state characterization of this system.

\section{Theoretical optimization of cat-state production}

\begin{figure}
\begin{centering}
\begin{tabular}{|c|c|}
\hline 
\tabularnewline
a.\includegraphics[scale=0.4]{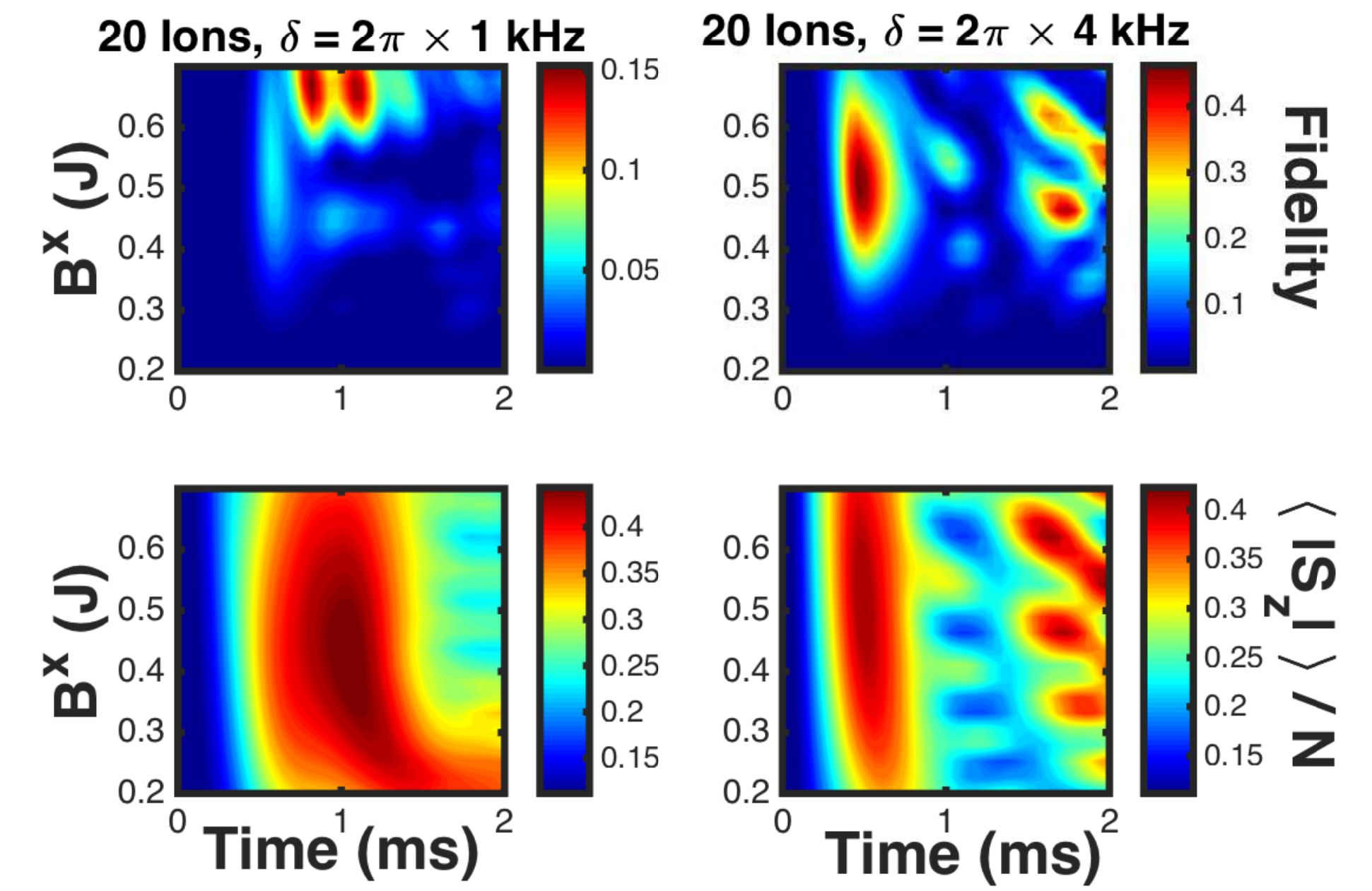} & b.\includegraphics[scale=0.4]{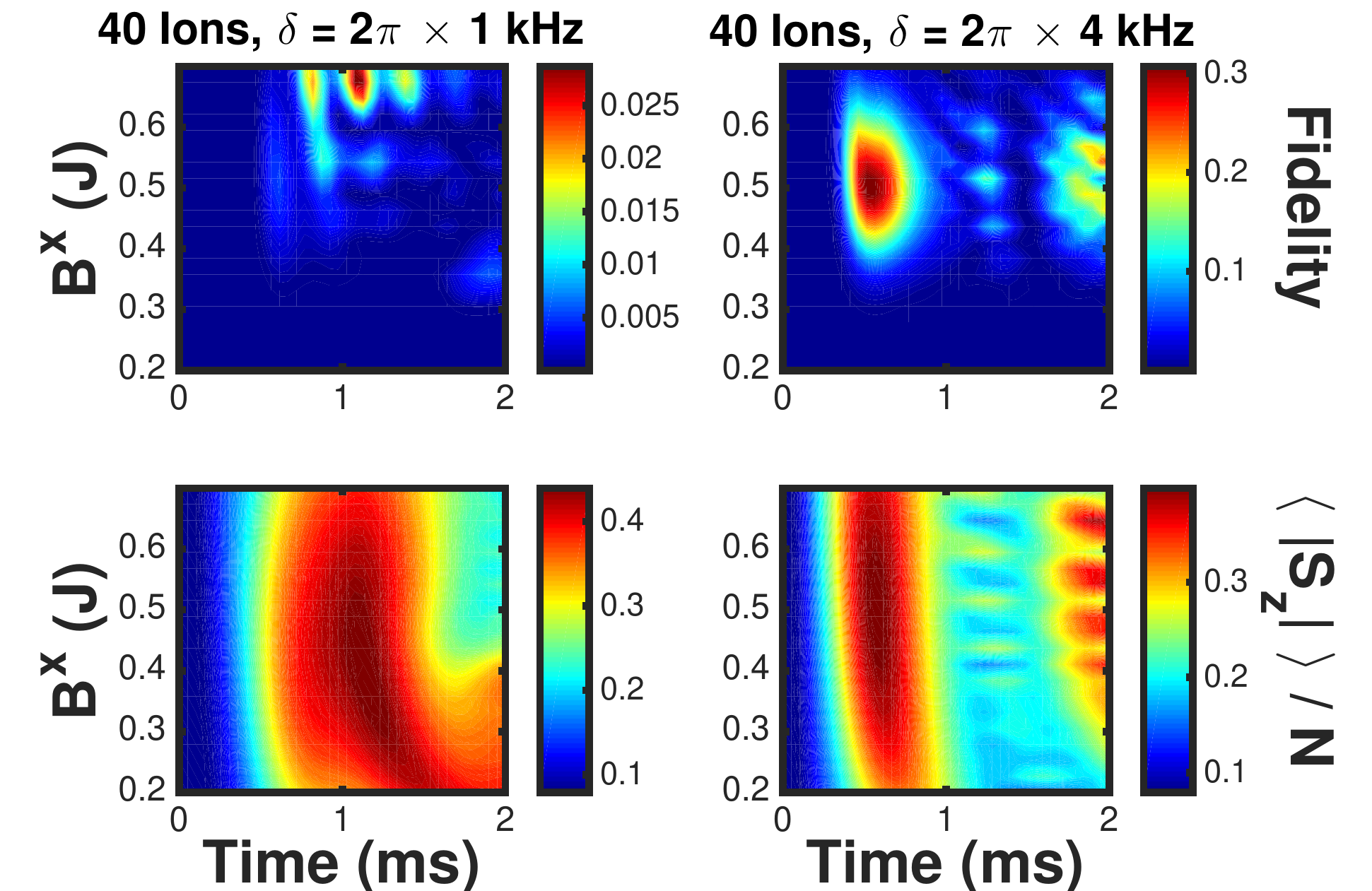}\tabularnewline
\hline 
\hline 
c.\includegraphics[scale=0.4]{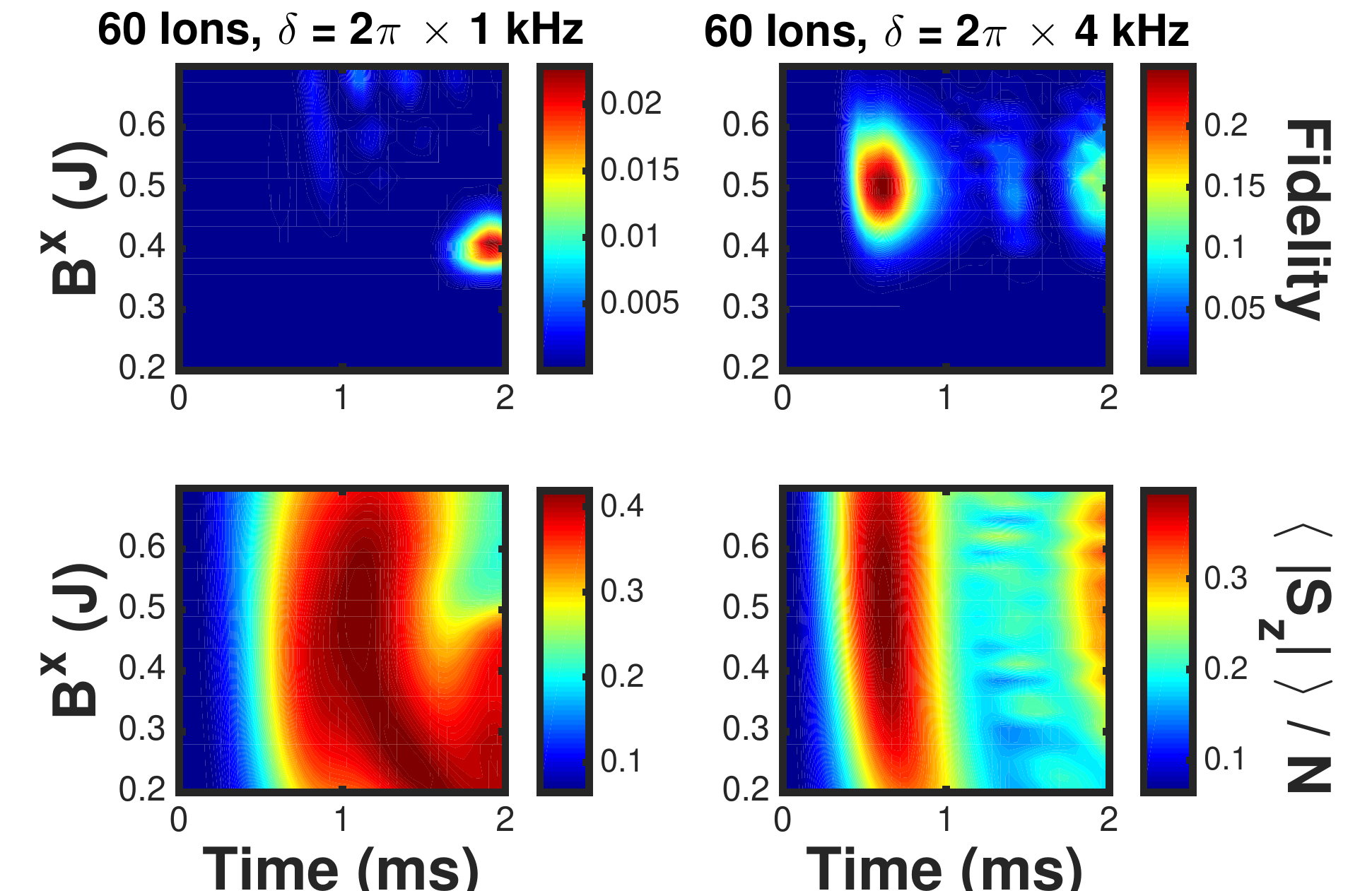} & d.\includegraphics[scale=0.4]{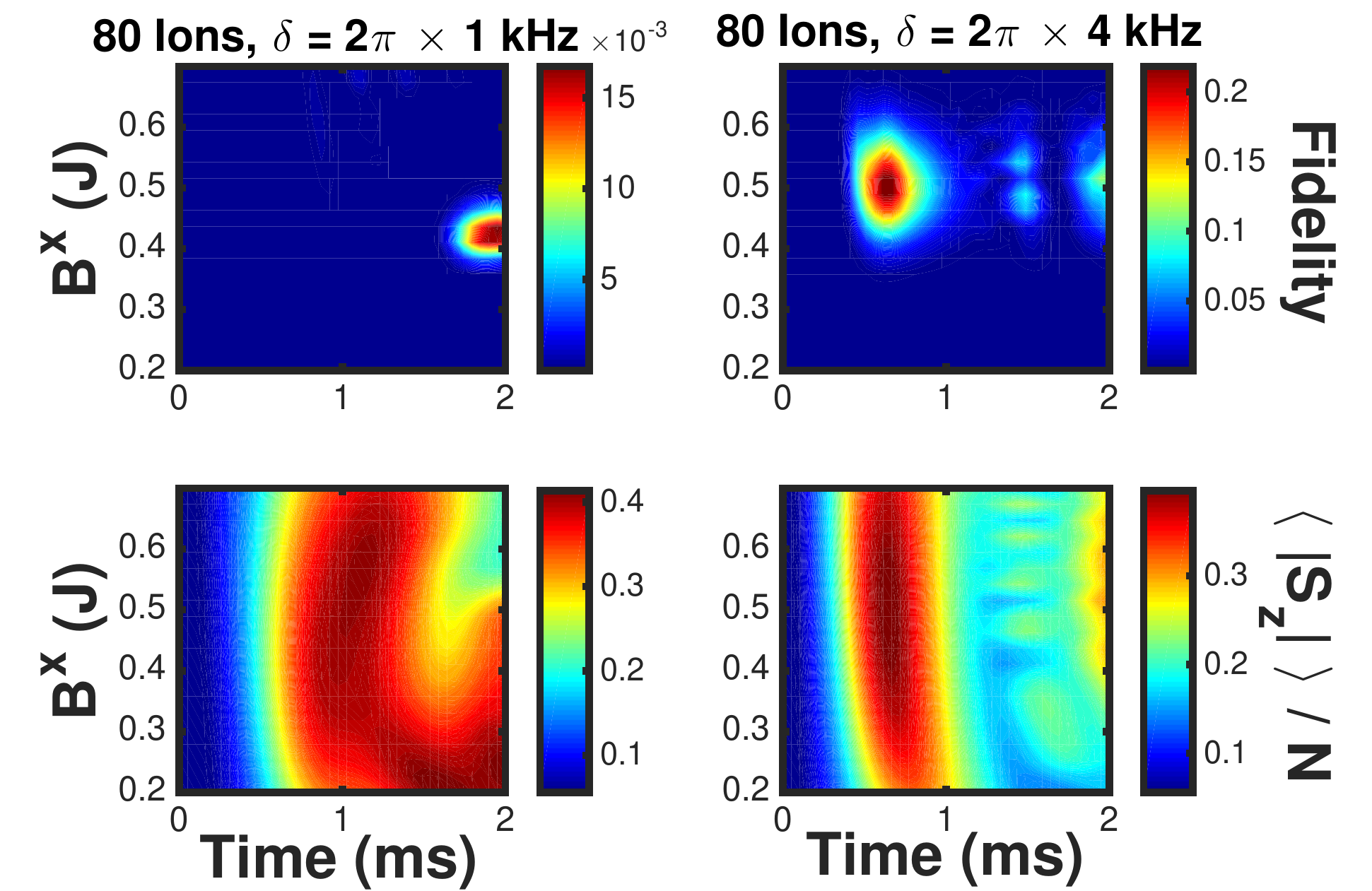}\tabularnewline
\hline 
\end{tabular}
\par\end{centering}

\caption{False color plots of the fidelity and the polarization for the theoretically calculated bang-bang protocol with system sizes
of 20, 40, 60 and 80 ions. These plots can be employed to optimize the bang-bang ramp profile. The top rows of each figure optimize the fidelity
of the Dicke ground state while the bottom rows optimize the value of $\langle|S_{z}/N|\rangle$.
Left columns of each figure are optimized using the near-critical detuning of $\delta=-2\pi\times1$~kHz
while the right columns are for  a detuning of $\delta=-2\pi\times4$~kHz.\label{fig: bang-bang}
}

\end{figure}

{\color{red} 
As discussed in the previous sections, and evidenced by the experimental data, a key challenge in the preparation of a cat-state is understanding the interplay between diabatic excitations and decoherence.
In simpler terms, mitigating diabatic excitations generically requires longer ramp times, but longer ramp times in turn magnify the effects of decoherence.
In this section, we follow the approach taken in~\cite{safavi2017exploring} and propose an ideal test case for the next generation of experiments.

We start by considering a detuning $\delta=-2\pi\times 4$~kHz, such that the spin-phonon resonance at $B^x \approx |\delta|$, is well-separated from the critical point at $B_c$.
This increases the size of the minimal energy gap, while the spins are still---to an excellent approximation---uniformly coupled to the COM mode.
Moreover, we assume that the initial thermal phonon occupation can be reduced to $\bar{n} \lesssim 0.2$, such that we can---to a good approximation---ignore this thermal contribution in the following calculations.
This parameter regime allows us to explore the potential of the bang-bang protocol, both for producing the ground-state, as well as for using it as a robust path to generating multi-partite entanglement.

In Fig.~\ref{fig: bang-bang}, we plot the preparation fidelity and the collective spin observable $\langle |S_z|/N\rangle$ for detunings of $\delta = -2\pi \times 1$~kHz and $\delta = -2\pi \times 4$~kHz and
for four different system sizes. The fidelity is calculated with respect to the ground-state of the Dicke model in the superradiant phase, and is given by
\begin{equation}
\mathcal{F}_{CAT}=|\langle \psi |CAT\rangle |^{2}.
\end{equation}
We find that for the larger magnitude $\delta=-2\pi\times 4$~kHz, the bang-bang shortcut performs best for ramp parameters $B^x\approx 0.5J$ and $t_{\rm hold}\approx 0.5$~ms.
This is evidenced by the maximal $\mathcal{F}_{CAT}$ as well as the peak in $\langle |\hat S_z|\rangle/N$. The fidelity ranges from $0.45$ for $20$ ions to $0.2$ for $80$ ions.
We highlight that this optimal ramp duration is short compared to the timescales on which decoherence has significant effect.

In contrast, for smaller magnitude detuning $\delta = -2\pi\times 1$~kHz, we do not find a significant correlation between the maximal fidelity and the maximal polarization of the spin.
In fact the maximal fidelity is only 0.14 for 20 ions and is as small as 0.016 for 80 ions, while the polarization remains large in the 0.4 range for all cases. We reconcile this observation
by noting that while diabatic excitations only slightly reduce the polarization ($m_{ex}<m_{CAT}=N/2$), they drastically reduce the ground-state fidelity since the excited states are orthogonal to the cat state.

In order to fairly evaluate the performance of the bang-bang protocol, we provide comparisons to the LA ramp. Guided by the previous calculations, we restrict to a system size of $20$ ions
and $\delta=-2\pi\times 4$~kHz where the LA ramp can produce rather large fidelities within $2$~ms.}

\begin{figure}
\begin{centering}
\begin{tabular}{|c|c|}
\hline 
a.\includegraphics[scale=0.4]{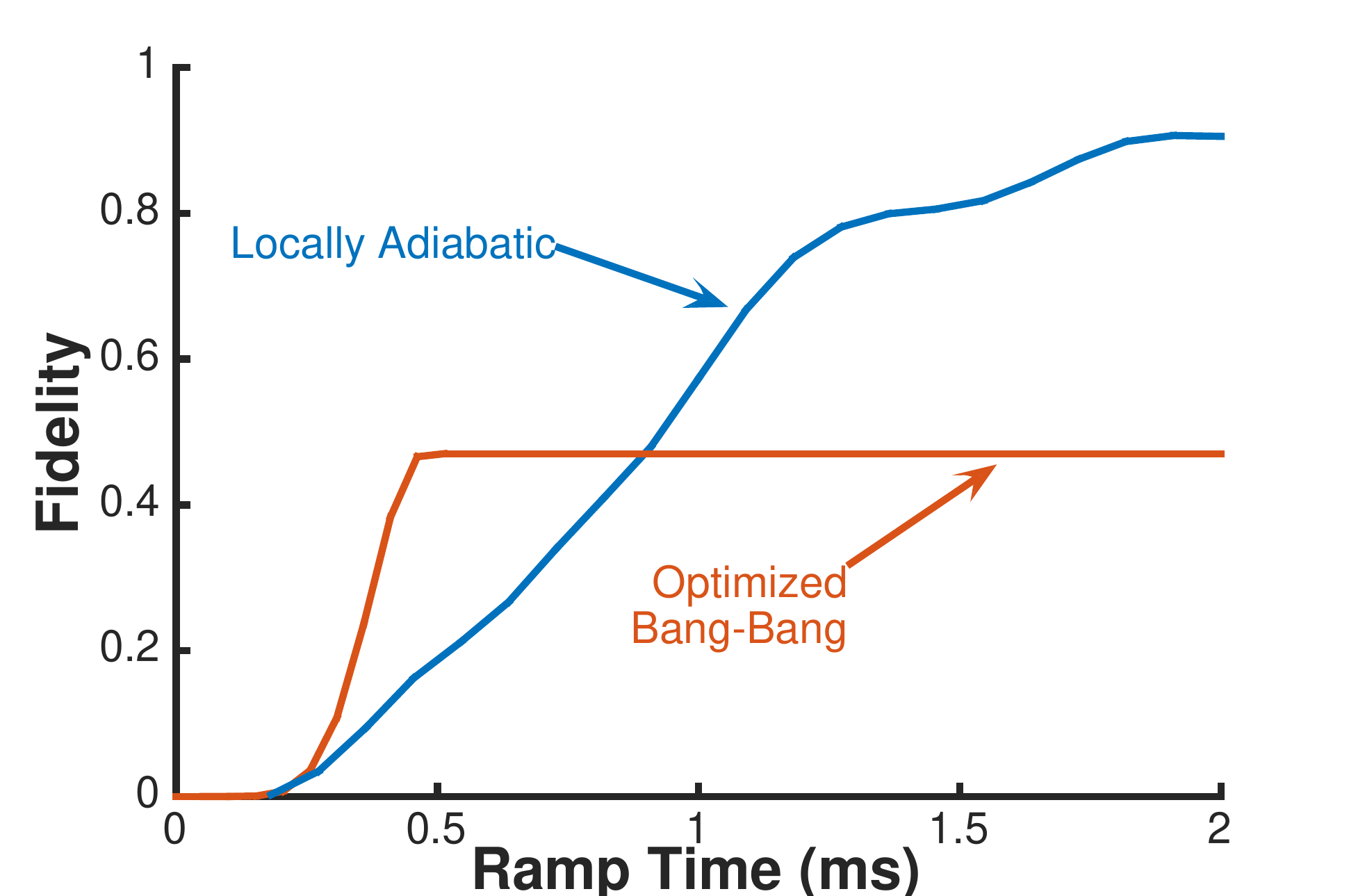} & b.\includegraphics[scale=0.4]{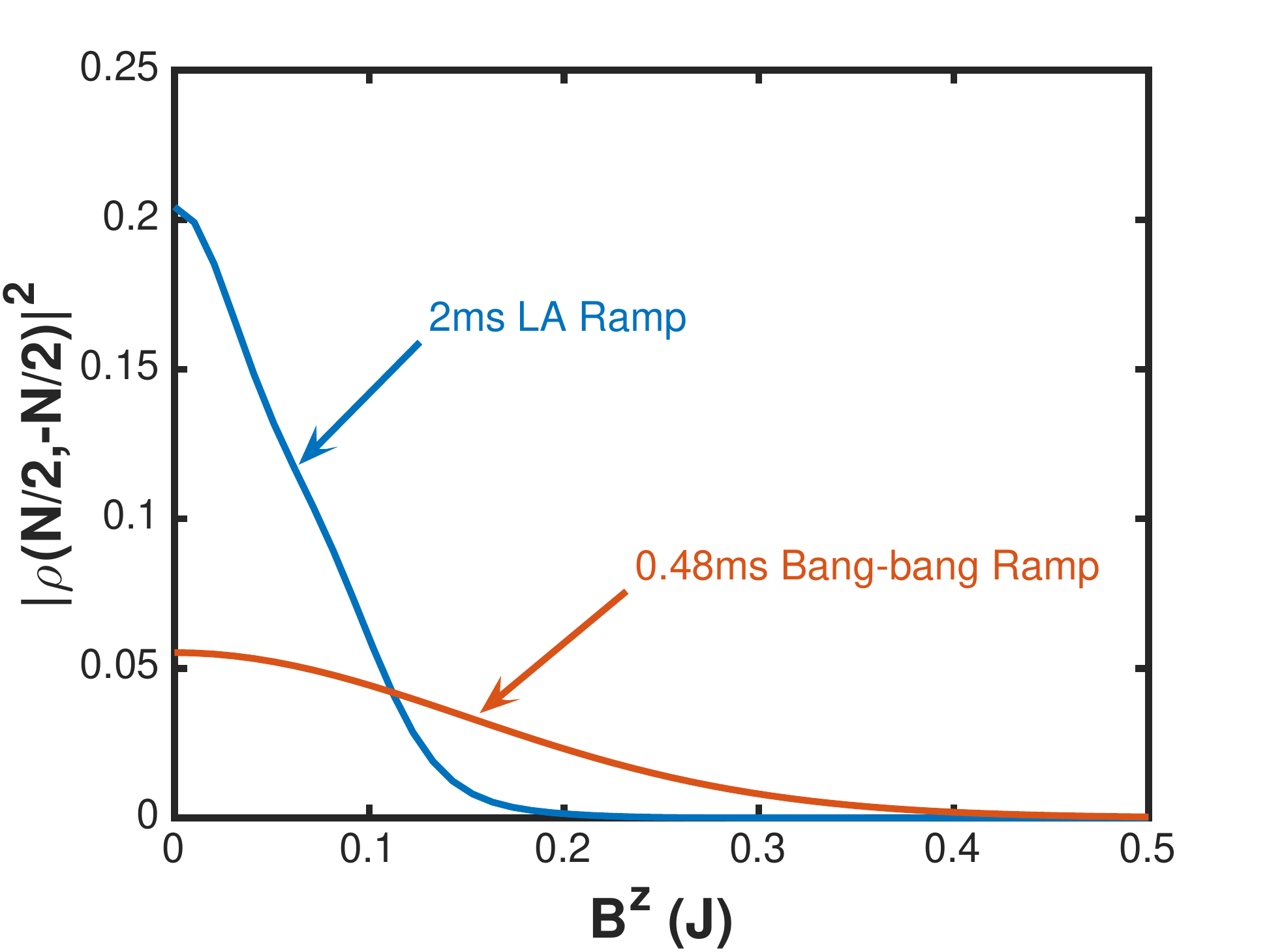}\tabularnewline
\hline 
\end{tabular}
\par\end{centering}

\caption{Theoretical predictions for a system of 20 ions with $J=2\pi\times 0.875$~kHz,
$\delta=-2\pi\times4$~kHz, and an initial phonon vacuum state. Panel
(a) shows the maximum ground-state fidelity as a function of ramp
time for both the bang-bang and LA ramps. The bang-bang approach outperforms the
LA ramp for times less than $0.9$~ms. Panel (b) shows the effects of adding a small longitudinal field on each ramping protocol for the case with 20 ions. 
The coherence of the cat state is obtained when a small longitudinal field is added
to the Dicke Hamiltonian. A 2 ms LA ramp is compared to an optimized bang-bang ramp of 0.485 ms. The slower decay in the bang-bang plot is due to the shorter ramp time.  \label{fig: long} }

\end{figure}

As shown in Fig.~\ref{fig: long}, the bang-bang shortcut always has a higher fidelity for $t<0.9$~ms. The LA ramp produces better fidelities for $t>0.9$~ms.  However, {\color{red} we note that when the maximal fidelity reached is $< 0.5$ it is insufficient to independently demonstrate non-trivial overlap with the entangled cat-state. Specifically, a fidelity of $0.5$ can
also be obtained with a statistical mixture of all spins up and all down. In the absence of decoherence, one may distinguish between the cat state and the maximally mixed state by measuring the
amplitude of the coherence} $|\rho_{N/2,-N/2}|= \mathcal{F}_{CAT}/2$. {\color{red} We note that, for a spin-phonon cat state, this coherence can be measured only after the disentangling procedure
discussed in Ref.~\cite{safavi2017exploring}. When significant decoherence is present, the verification of cat state coherence requires full characterization of the state.}

\begin{figure}
	\begin{centering}
		\begin{tabular}{|c|c|}
			\hline 
			a.\includegraphics[scale=0.4]{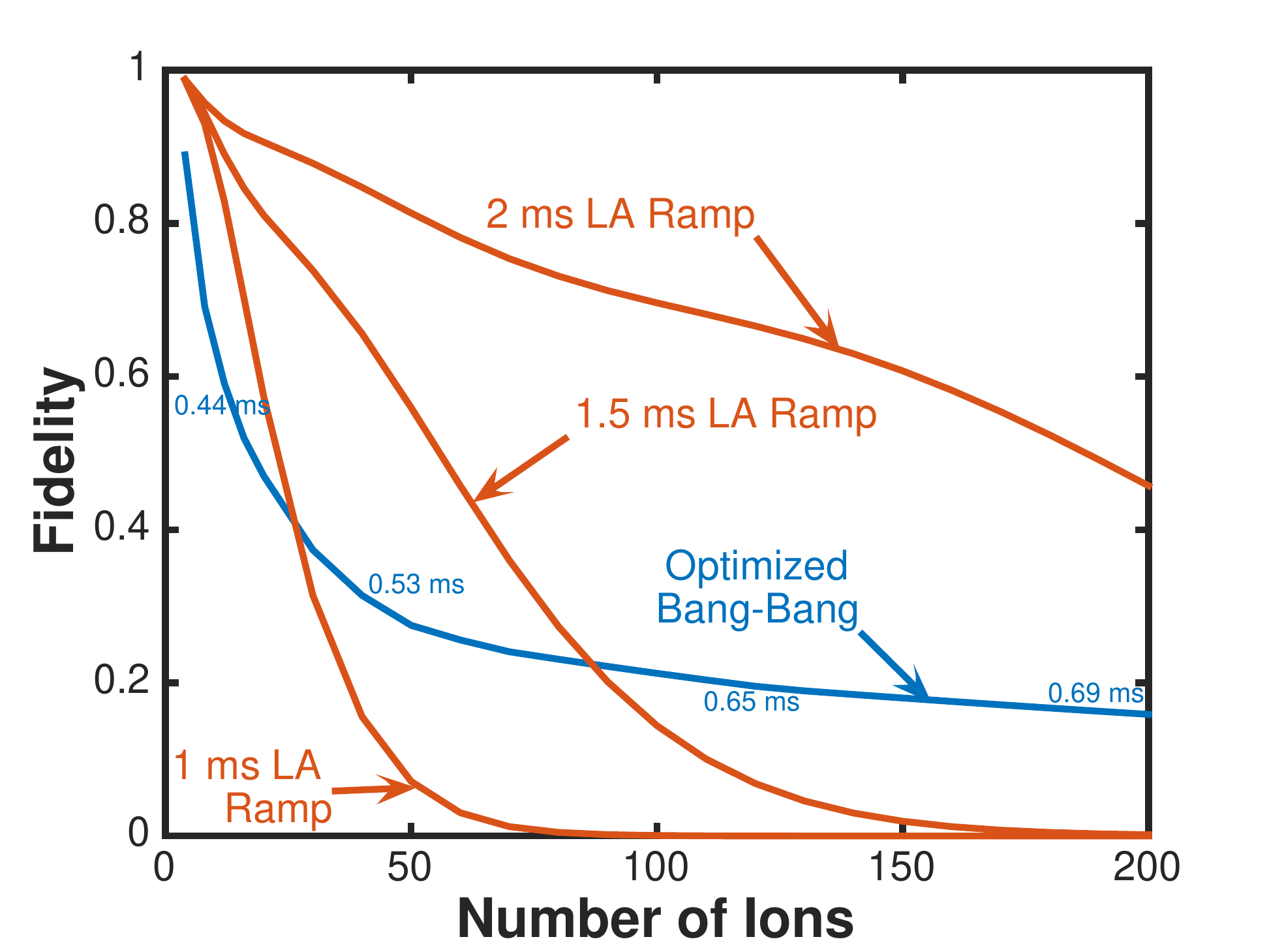} & b.\includegraphics[scale=0.4]{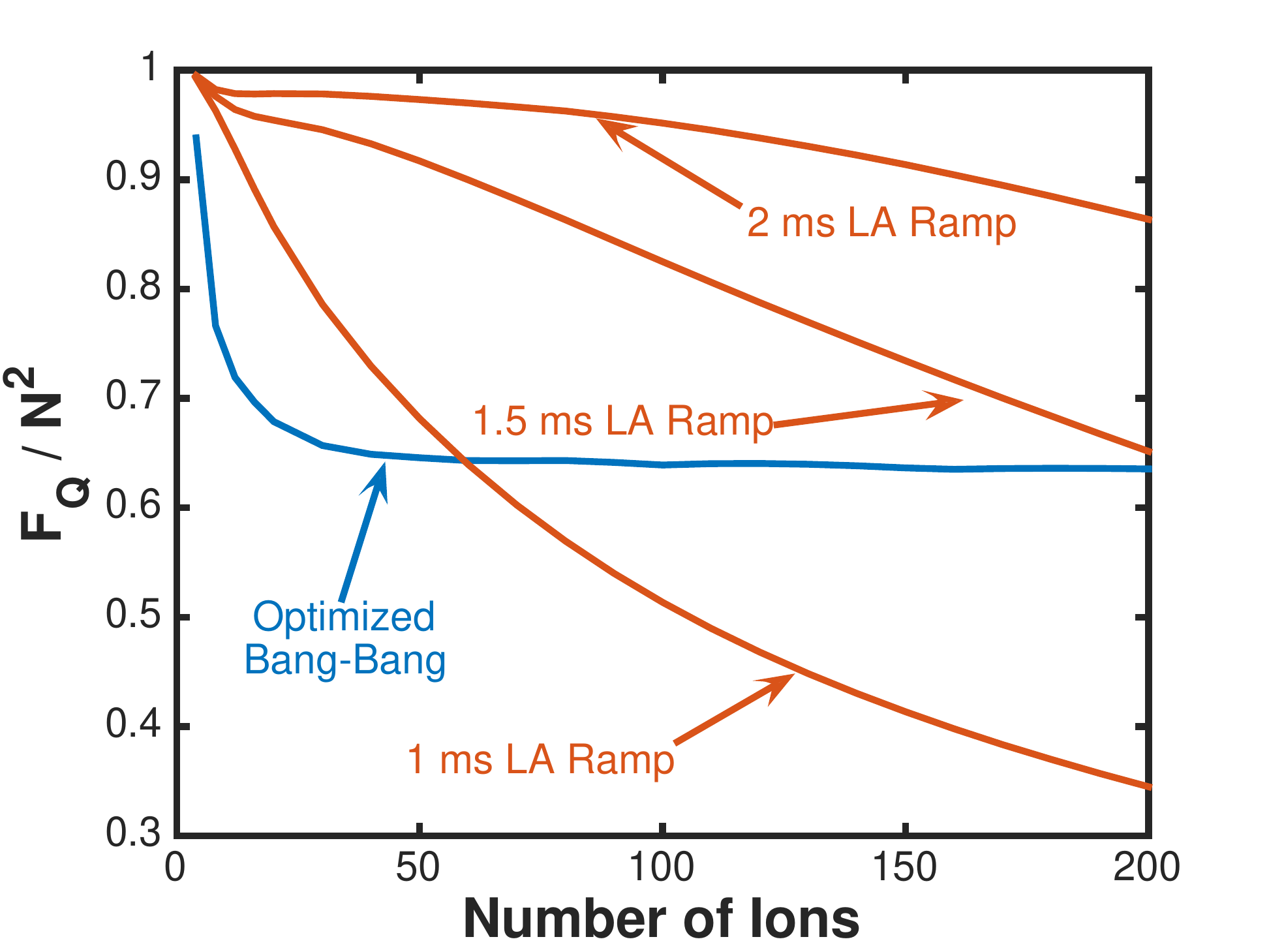}\tabularnewline
			\hline 
		\end{tabular}
		\par\end{centering}
	
	\caption{Panel (a) shows how the ground-state fidelity of the Dicke
		Hamiltonian scales as a function of system size for bang-bang and
		LA ramps of 1.0, 1.5 and 2.0~ms. The bang-bang approach outperforms the LA
		ramps of 1.5~ms at 90 ions. Panel (b) shows how the multiparticle
		entanglement depth scales as a function of system size. \label{fig: size} }

\end{figure}

{\color{red} In Fig.~\ref{fig: size}, we show the scaling of the ground-state fidelity with system size. We find that, for fixed ramp times, both protocols perform worse as the system size
increases. However the bang-bang protocol appears to be less sensitive to increasing system size for shorter ramp times.

So far  we have considered idealized conditions for the ramping protocols. However, a common experimental imperfection to consider in a Penning trap is a residual longitudinal field, which can break the
degeneracy of the ground-state and thus degrade the preparation of the cat-state. In Fig.~\ref{fig: long}, we illustrate the effect of a fixed uniform longitudinal field, $\propto B^z \hat{S}^z$, on the coherence of
the spin-phonon cat state.
Here, the coherence of the spin-phonon cat state is defined as $\langle N/2|\langle \alpha| \rho |-\alpha \rangle |-N/2\rangle$. As the spin-phonon reflection parity is
no longer a symmetry of the model, the initially purely odd or even parity ground-states begin to mix as the state is evolved forward in time. In this example, the state is initialized in the
even parity manifold. One can see that the longitudinal field causes the final coherences to decay with the effect being more dramatic for the LA ramp than for the bang-bang ramp.
The slower decay of the spin-phonon cat state coherence is a result of the bang-bang ramp having a shorter ramp time \cite{safavi2017exploring}.}

{\color{red}
Our discussion up to now has focused on the ground-state fidelity to characterize the performance of each protocol. However, an equally important measure is the metrological useful entanglement, which
we quantify using the quantum Fisher information (QFI).
For a pure state the Fisher information is given by
\begin{equation}
F_{Q}=4\langle (\Delta \hat{S}_{\vec n} )^2 \rangle,
\label{eq: fisher}
\end{equation}
where $\Delta(\hat{S}_{\vec{n}})^{2}$ is the variance of $\hat{S}_{\vec{n}}$, and $\vec{n}$ the spin direction that maximizes the QFI~\cite{braunstein1994statistical}. The QFI is an effective witness of multipartite entanglement in the following sense: $F_{Q}\geq N$ implies that entanglement is present in the system and full N-body entanglement is classified as $F_{Q}\geq N^{2}/2$. Maximal entanglement, in this context, refers to a saturation of the bound for the quantum Fisher information, $F_{Q}=N^{2}$, which represents the result for the spin-phonon cat
state. In Fig.~\ref{fig: size}(b), we show the behavior of QFI as a function of system size. We find that QFI is much less sensitive to the size of the system, implying that the number of diabatic excitations do not degrade the QFI as severely as the ground-state fidelity.
This occurs, in part, because we are symmetry restricted to the spin multiplet, which when $B^{x}\rightarrow0$ exhibits N-partite entanglement for every eigenstate, due to the parity symmetry. It is interesting to note that while the bang-bang entanglement quickly drops off at small $N$, it appears to approach a constant
value of $0.65N^{2}$, which is still a quite large entanglement depth for
systems on the order of hundreds of ions, and for short ramp times. We don't know why the bang-bang ramp approaches this limit.

Finally, we note that both the fidelity and QFI  will be affected by decoherence processes as discussed in the previous section. While modeling the exact effect of decoherence is beyond the scope of this work, it is expected that the impact will scale with the ramp time. As such, while the bang-bang protocol may not be able to create experimentally useful fidelities, it will be a valuable approach for generating large QFI even in the presence of appreciable decoherence. Clearly faster protocols are more resilient to decoherence. Furthermore, since the effects of decoherence are magnified as the number of ions increases, the bang-bang protocol may provide a robust, and experimentally feasible path to creating states for quantum enhanced metrology.}

\section{Conclusions}

We have shown that the bang-bang protocol as applied to the Dicke
model can be easily realized in Penning trap quantum simulators. This
shortcut to adiabaticity is clearly superior to the alternative LA approach in terms of the creation of metrologically useful entangled states on short time-scales. The bang-bang approach also scales
better with larger system sizes when compared to the LA ramp. The ability to generate entanglement rapidly for large systems has crucial implications for future experiments, where decoherence is a key consideration. Specifically the bang-bang protocol has the potential to easily create highly entangled states of hundreds or even thousands of ions. 

\section{Acknowledgments}
The authors acknowledge fruitful discussions with J.
Marino, M. Holland and K. Lehnert. A. M. R
acknowledges support from Defense Advanced Research
Projects Agency (DARPA) and Army Research Office grant W911NF-16-1-0576, NSF grant PHY-1521080,
JILA-NSF grant PFC-173400, and the Air Force Office
of Scientific Research and its Multidisciplinary University
Research Initiative grant FA9550-13-1-0086. M.G.
acknowledges support from the DFG Collaborative Research
Center SFB1225 (ISOQUANT). E. J. also acknowledges
support from Leopoldina Fellowship Programme.
JKF and JC acknowledge support from NSF
grant PHYS-1620555. In addition, JKF acknowledges
support from the McDevitt bequest at Georgetown. Financial
support from NIST is also acknowledged.

\section*{References}
\bibliographystyle{iopart-num}

\bibliography{bang_iop}

\end{document}